# Terahertz Fourier Ptychographic Imaging

Pitambar Mukherjee[1], Vivek Kumar[1,2], Frederic Fauquet[1], Amaury Badon[3], Damien Bigourd[1], Kedar Khare[4], Sylvain Gigan[2], Patrick Mounaix[1]

[1] IMS Laboratory, University of Bordeaux, UMR CNRS 5218, 351 Cours de la Libération Bâtiment A31, 33405 Talence, France.
[2] Laboratoire Kastler Brossel, ENS-Université PSL, CNRS, Sorbonne Université, Collège de France, 24 rue Lhomond, 75005 Paris, France.
[3] Laboratoire Photonique Numérique et Nanosciences (LP2N), UMR 5298, University of Bordeaux, 33400 Talence, France.
[4] Optics and Photonics Centre, Indian Institute of Technology Delhi, New Delhi 110016, India.

**High-resolution imaging in the terahertz (THz) spectral range remains fundamentally constrained by the limited numerical apertures of currently existing state-of-the-art imagers, which restricts its applicability across many fields, such as imaging in complex media or non-destructive testing. To address this challenge, we introduce a proof-of-concept implementation of THz Fourier Ptychographic imaging to enhance spatial resolution without requiring extensive hardware modifications. Our method employs a motorized kinematic mirror to generate a sequence of controlled, multi-angle plane-wave illuminations, with each resulting oblique-illumination intensity image encoding a limited portion of the spatial-frequency content of the target imaging sample. These measurements are combined in the Fourier domain using an aberration-corrected iterative phase-retrieval algorithm integrated with an efficient illumination calibration scheme, which enables the reconstruction of resolution-enhanced amplitude and phase images through the synthetic expansion of the effective numerical aperture. Our work establishes a robust framework for high-resolution THz imaging and paves the way for a wide array of applications in materials characterization, spectroscopy, and non-destructive evaluation.**

## 1. INTRODUCTION

THz radiation occupies a scientifically fertile yet technologically underdeveloped region of the electromagnetic spectrum. Its exceptional ability to penetrate a wide range of non-conductive and weakly absorbing materials has positioned the THz band as a promising frontier for non-destructive imaging across diverse applications, including material inspection [1–4], biomedical diagnostics [5,6], cultural heritage conservation [7–9], and security screening [10,11]. While the potential breadth of THz applications is increasingly acknowledged, the practical realization of high-quality THz imaging remains hampered by significant physical and technological constraints. Unlike optical imaging, which benefits from mature detector arrays, high numerical aperture (NA) optics, and a century of algorithmic and fabrication refinements, THz imaging contends with long wavelengths, limited spatially coherent sources, low detector sensitivity, and a persistently limited space-bandwidth product (SBP) [12]. These constraints collectively lead to low resolution, limited contrast, and an absence of robust phase-reconstruction frameworks, substantially reducing the practical impact of THz imaging systems. To address challenges associated with THz imaging, techniques such as near-field [13–15], computational Ghost imaging [16–19], and synthetic aperture [20] have been sensibly diffused with the state-of-the-art time-domain spectroscopy (TDS) photonic framework. Although such integration offers coherent pulsed THz imaging over the broad THz spectral range, however, TDS technique fundamentally relies on femtosecond laser pulses, which introduce significant system complexity and cost [21].

Recent advancements in photonic continuous-wave (CW) THz systems have emerged as an alternative approach for implementing and exploring imaging capabilities by adopting cutting-edge quantum cascade lasers (QCLs) and microbolometric-based matrix detectors. Such a CW platform notably offers technical benefits in reducing the signal acquisition time by eliminating the need for synchronized pulses and delay lines, which are typically required in TDS-based imaging systems. Subsequently, researchers have employed CW-THz platforms to develop efficient imaging techniques, where the challenges shift towards computational recovery of the complex spatial field profile of the target sample. In this context, various techniques such as interferometric and non-interferometric imaging methods [22–25], multi-plane phase retrieval [26–29], diffraction tomography [30–32], along with the integration of deep learning approaches [33,34] have been explored. Such noticeable advances have significantly broadened the operational scope of THz imaging. However, the performance of traditional THz imaging has long been constrained by the SBP, which represents a fundamental trade-off between field-of-view (FOV) and spatial resolution, both of which are related to the spatial bandwidth available in the THz imaging system. For instance, at 1 THz ($\lambda = 0.3$ mm), the far-field image resolution is limited to $\sim 180$ μm, rendering high-resolution THz imaging fundamentally unattainable solely through physical aperture [12]. In this context, it is worth noting that increasing the SBP in a conventional THz imaging system will inevitably introduce system design complexity, making it difficult to minimize the optical aberrations.

Interestingly, to overcome the fundamental limitations posed by the trade-off between spatial resolution and FOV in diffraction-limited imaging systems, Fourier Ptychography has emerged as a viable computational imaging technique, gaining significant attention within the optical spectral range [35–37]. In essence, the Fourier Ptychography resolution-enhancement mechanism arises from digitally processing a sequence of transmitted diffraction-limited images through the sample, acquired by illuminating the sample with waves at various incident angles. Notably, when an oblique plane wave illuminates a sample in a coherent imaging framework, it introduces an additional linear phase into the field at the sample plane. This results in a shift of the spatial spectrum of the scattered field, allowing the capture of higher spatial frequency information by a lens with a low NA. By combining images taken under various illumination angles, which contains limited spatial-frequency components of the sample, one can numerically synthesize an image with frequency support well beyond the intrinsic limit set by the physical NA of the lens [38,39].

In the context of accessing newly emerging spectral domains, i.e., THz, Fourier Ptychography presents both the enticing possibility of closing long-standing technological gaps and, simultaneously, a new landscape of practical challenges. For instance, illumination engineering in the THz frequency range remains relatively underdeveloped compared to optics [40]. While optical Fourier Ptychography benefits from integration of LED arrays and high-speed spatial light modulators for the source beam modulation, analogous devices are not readily available at THz frequencies. Instead, a THz Fourier Ptychography platform could employ mechanically actuated beam-tilting elements. However, such optical components are susceptible to thermal drift and vibrational instabilities, which complicates the precise control of illumination angle needed for reliable synthetic-aperture acquisition. Therefore, achieving stable and repeatable multi-angle measurements with a QCL exhibiting partial spatial coherence is itself a nontrivial challenge, demanding careful calibration and compensation strategies tailored to the dynamics and sensitivity of THz hardware. Moreover, because the longer operating wavelength in THz imaging systems undergoes stronger diffraction, the captured signal inherently lacks higher spatial frequency components [41]. Such a limitation necessitates the larger NA collection optics, i.e., a large area lens. However, common THz lens materials exhibit significant refractive and absorptive dispersion, which constrains feasible optical designs and degrades achievable imaging performance. Compounding to this, commercially available detector arrays typically offer coarse pixel pitches (for example, current state-of-the-art microbolometric THz cameras offer 35 μm pixel pitches [42]), which are significantly smaller than the operative wavelength, resulting in additional discrepancies in sampling and resolution that are not encountered in optical Fourier Ptychography. Consequently, one could argue that the fundamental operational deviations encountered in adapting optical Fourier Ptychography to the THz spectral range represent not merely a trivial reimplementation, but rather a substantial conceptual and technical reformulation of the Fourier Ptychography framework.

Building upon this, we introduce in this work a THz Fourier Ptychographic imaging technique based on multi-angle plane-wave illumination. We employ a motorized kinematic mirror to produce a sequence of oblique THz illuminations, each shifting different regions of the sample's spatial frequency spectrum into the passband of a low-NA detection system. The collected diffraction-limited intensity measurements, each capturing a distinct Fourier patch, are computationally integrated to synthesize a large effective NA, significantly surpassing that of the physical optics, while preserving the full FOV. Our imaging framework incorporates an aberration-corrected iterative phase retrieval algorithm that relies on precise knowledge of the illumination produced by the motorized kinematic mirror. To this end, we implement a circular edge-detection-based calibration routine to estimate the wave vector of the impinging beam, eliminating geometric misalignment errors and mitigating the effects of thermal drift and mechanical instabilities. Utilizing an illumination calibration mechanism combined with an iterative phase retrieval routine, we reconstruct high-resolution amplitude and phase profiles of the sample, demonstrating quantitative THz phase imaging capabilities that are previously unattainable with conventional THz imaging systems.

## 2. METHODOLOGY

We present a computational imaging approach based on THz Fourier Ptychography to overcome the diffraction-limit inherent in conventional imaging systems. A conceptual overview of our imaging scheme is illustrated in Figure 1(a), where a target imaging sample is sequentially probed by a series of plane waves incident at several angles. Under this scenario, a thin semi-transparent object $O(x, y)$ is illuminated by a plane wave described as $exp(i(k_{xn}x + k_{yn}y))$, where the incident wave vector is defined as $\vec{k}_n \equiv (k_{xn}, k_{yn})$. In the experimental realization, a THz beam emitted from a CW-QCL operating at 3.5 THz is first collimated by an optical assembly L1. The incident wave vector of the plane wave is then precisely controlled using a motorized tilting mirror and projected onto the sample via a combination of two off-axis parabolic mirrors (P1 and P2). Further, the transmitted scattered wave intensity through the sample is recorded using a microbolometer THz camera placed at the image plane of a 4f telescopic lens arrangement L2 and L3. Under these conditions, the recorded intensity image at the detector can be modeled as the convolution of the scattered field emerging from the sample, and the spatially invariant point spread function $p(x, y)$ of the 4f imaging system, expressed as:

$$I^{(n)}(x, y) = \left|O(x, y) \, exp\left(i(k_{xn}x + k_{yn}y)\right) \otimes p(x, y)\right|^2 \quad (1)$$

where $\otimes$ denotes the convolution operation. To process these images in the spatial Fourier domain, the intensity image spectrum can be obtained as,

$$\tilde{I}^{(n)}(k_x, k_y) = \tilde{O}(k_x - k_{xn}, k_y - k_{yn})\tilde{p}(k_x, k_y) \\ \star \tilde{O}(k_x - k_{xn}, k_y - k_{yn})\tilde{p}(k_x, k_y) \quad (2)$$

where $\star$ represents autocorrelation, $\tilde{O}(k_x, k_y) = \mathcal{F}_{2D}[O(x, y)]$ is the spatial Fourier transform of the sample, $\tilde{p}(k_x, k_y) = \mathcal{F}_{2D}[p(x, y)]$ is the pupil function of the imaging system and $\mathcal{F}_{2D}$ is a spatial Fourier transform operator. The term $\tilde{O}(k_x - k_{xn}, k_y - k_{yn})\tilde{p}(k_x, k_y)$ corresponds to a shifted spectrum of the object with the circle $k_x^2 + k_y^2 \leq \left(\frac{NA}{\lambda}\right)^2$ and zero elsewhere. Physically, the autocorrelation operation scans two copies of the shifted and pupil-filtered object spectrum across each other, coherently summing the spectral components at each frequency offset to give $\tilde{I}^{(n)}(k_x, k_y)$. This concept is illustrated in Figure 1(b), where we show diffraction-limited intensity images and corresponding spatial Fourier transforms recorded under three different illumination wave vectors $\vec{k}_1$, $\vec{k}_2$, and $\vec{k}_3$, following the spiral illumination trajectory in illumination k-space. This implies that the offset in the wave vector $\vec{k}_n \equiv (k_{xn}, k_{yn})$ for recorded images directly correlates with the illumination angle of the probing beam. Consequently, tweaking the illumination angle is equivalent to shifting the center of the pupil aperture. Therefore, trivially in this vision, by capturing multiple diffraction-limited images corresponding to a series of illumination wave vectors, each containing a limited spatial frequency spectrum, and coherently combining them, one can effectively synthesize an expanded spatial frequency passband beyond the conventional diffraction limit. Following this, a sequence of diffraction-limited, low-resolution intensity images is acquired under varying illumination wave vectors of the probing beam, represented as $\vec{k}_n$, where $n \in [1\ N]$ and N is the total number of illuminations. These measurements are then computationally processed using an aberration-corrected iterative reconstruction algorithm designed to recover both the pupil function and complex spatial field of the sample that satisfy equations 1 and 2 for all measured intensity images. In instances where a precise estimate of the pupil function is accessible from prior characterization of the imaging system, a conventional iterative phase retrieval algorithm [38] can be applied to directly determine $\tilde{O}(k_x, k_y)$, satisfying equation 1. However, since such algorithms update only the sample spectrum while keeping the pupil function unchanged, inaccuracies in the estimated pupil function can degrade the reconstruction quality. Such inaccuracies often stem from insufficient aberration modeling during pre-characterization or from mechanical or optical drifts within the imaging setup. A similar challenge has been encountered in conventional Ptychography [43], where the probe illumination is spatially scanned across the sample and the far-field diffraction patterns are recorded. These recorded diffraction patterns are subsequently processed using a traditional ptychographic iterative engine, which relies on an accurate knowledge of the probe function to retrieve the phase of the target sample [44]. However, in practice, the probe function is often imperfectly known due to uncertainties in the aperture or focusing optics that shape the illumination beam. To address this limitation, several approaches have been proposed in the optical frequency range that jointly optimize both the object and the probe functions [45–47]. Drawing inspiration from the dual reconstruction technique, particularly by the extended Ptychographic iterative engine framework [48,49], we adopt a similar integrated strategy to compensate for optical aberrations inherent to the THz imaging systems. However, in these Ptychography approaches, a digital micromirror device (DMD) or an LED array is employed to induce tilt in the incident beam to effectively probe the sample.

Our computational imaging routine begins with assuming initial estimates for the spatial spectrum of sample and pupil function, $\tilde{O}^{(0)}(k_x, k_y)$ and $\tilde{p}^{(0)}(k_x, k_y)$, respectively. By using the $n^{th}$ raw intensity image, the updated spatial spectrum of the sample $\tilde{O}^{(n)}(k_x, k_y)$ and the pupil function $\tilde{p}^{(n)}(k_x, k_y)$ can be obtained as follows. First, we simulate a low-resolution field image $\xi_l^{(n)}(x,y)$ limited by the NA of the 4f imaging lens system, utilizing the non-updated sample spectrum, and expressed as,

$$\xi_l^{(n)}(x,y) = \mathcal{F}_{2D}^{-1}\{\tilde{O}^{(n-1)}(k_x - k_{x(n-1)}, k_y - k_{y(n-1)})\tilde{p}^{(n-1)}(k_x, k_y)\} \quad (3)$$

where, $\tilde{O}^{(n-1)}$, $\tilde{p}^{(n-1)}$ represents the non-updated spatial spectrum of sample and pupil function, respectively, and $\mathcal{F}_{2D}^{-1}$ is the inverse spatial Fourier transform operation. In this scenario, the illumination is modeled as an oblique plane wave with an impinging wavevector $\vec{k}_{(n-1)} \equiv (k_{x(n-1)}, k_{y(n-1)})$, which introduces a corresponding spatial frequency shift of $(k_{x(n-1)}, k_{y(n-1)})$. Next, the simulated field is updated by imposing the intensity constraint, wherein the amplitude of the simulated field distribution is replaced by the square root of the measured intensity image $I^{(n)}(x,y)$:

$$\xi_{updated}^{(n)}(x,y) = \sqrt{I^{(n)}(x,y)} \frac{\xi_l^{(n)}(x,y)}{|\xi_l^{(n)}(x,y)|} \quad (4)$$

and obtained the corresponding spatial spectrum of the updated low-resolution diffraction-limited field as,

$$\tilde{\xi}_{updated}^{(n)}(k_x, k_y) = \mathcal{F}_{2D}\{\xi_{updated}^{(n)}(x,y)\} \quad (5)$$

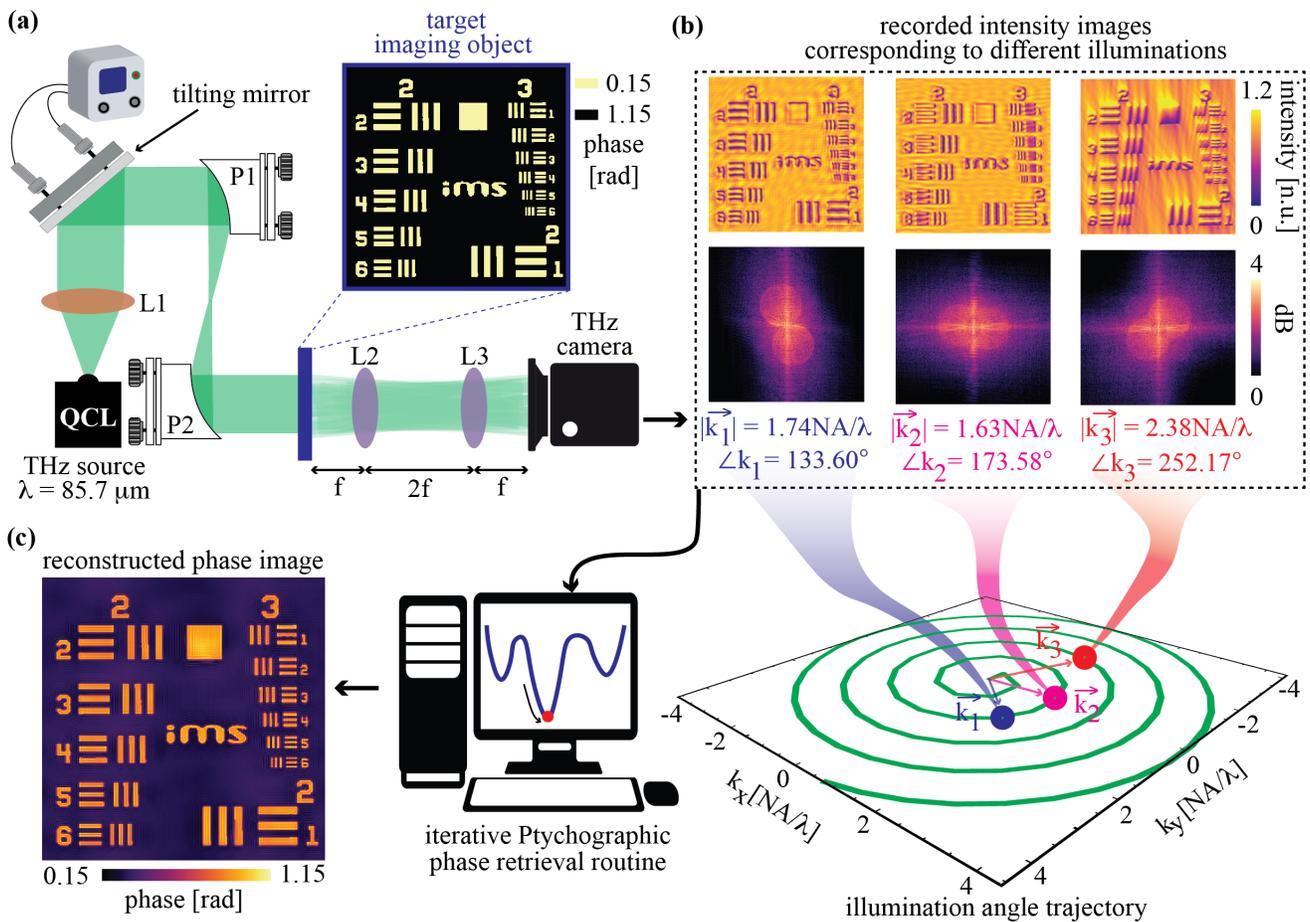

**Figure 1: Operative design of THz Fourier Ptychographic imaging. (a)** Schematic illustration of the methodology for multi-angle illumination-based THz phase imaging. A collimated THz beam is directed toward the sample under a sequence of controlled illumination angles using a motorized tilting mirror. The corresponding diffraction-limited intensity images are recorded by the THz camera, and an iterative Ptychographic phase retrieval routine enables the reconstruction of a high-resolution phase image. L1: beam collimating optics, P1 and P2: off-axis parabolic mirrors, L2 and L3: lenses forming the 4f imaging system. **(b)** Simulated diffraction-limited images obtained at different illumination angles, along with their corresponding spatial Fourier-transformed and the spiral illumination trajectory in Fourier space. **(c)** Reconstructed phase image of the imaging object obtained by computationally processing all the diffraction-limited images acquired under diverse illumination angles.

Using this updated spectrum, both the pupil function and the sample spectrum can be obtained as,

$$\tilde{p}^{(n)}(\vec{k}) = \tilde{p}^{(n-1)}(\vec{k}) + \alpha(\vec{k}) \quad (6)$$
$$\tilde{O}^{(n)}(\vec{k}) = \tilde{O}^{(n-1)}(\vec{k}) + \beta(\vec{k}) \quad (7)$$

Here, we obtain $\alpha(\vec{k}) = \frac{|\tilde{O}^{(n-1)}(\vec{k}-\vec{k}_n)|[\tilde{O}^{(n-1)}(\vec{k}-\vec{k}_n)]^*[\tilde{\xi}_{updated}^{(n)}(\vec{k}) - \tilde{O}^{(n-1)}(\vec{k}-\vec{k}_n)\tilde{p}^{(n-1)}(\vec{k})]}{|\tilde{O}^{(n-1)}(\vec{k})|_{max}(|\tilde{O}^{(n-1)}(\vec{k}-\vec{k}_n)|^2 + \sigma_1)}$

and $\beta(\vec{k}) = \frac{|\tilde{p}^{(n-1)}(\vec{k}+\vec{k}_n)|[\tilde{p}^{(n-1)}(\vec{k}+\vec{k}_n)]^*[\tilde{\xi}_{updated}^{(n)}(\vec{k}+\vec{k}_n) - \tilde{p}^{(n-1)}(\vec{k}+\vec{k}_n)\tilde{O}^{(n-1)}(\vec{k})]}{|\tilde{p}^{(n-1)}(\vec{k})|_{max}(|\tilde{p}^{(n-1)}(\vec{k}+\vec{k}_n)|^2 + \sigma_2)}$

where * represents the complex conjugate, and $\sigma_1$ and $\sigma_2$ are small regularization constants introduced to prevent numerical instabilities caused by near-zero denominators. Additionally, without loss of generality, we represent wavevectors as $\vec{k} \equiv (k_x, k_y)$ and $\vec{k}_n \equiv (k_{xn}, k_{yn})$ in the above equations. Since for a coherent imaging system, the pupil function is a circular low-pass spatial frequency filter with diameter $\frac{NA}{\lambda}$, we set the values outside the circular region to zero during iterative reconstruction. This process is repeated across all the N recorded intensity frames, $I^{(n)}(x,y)$, completing one full cycle of pupil and object spectrum updates. The overall routine is then iterated multiple times until convergence, achieving the final estimates of the high-resolution amplitude and phase of the sample. Conversely, the reconstructed high-resolution complex field distribution is obtained by performing an inverse spatial Fourier transform of the final recovered spectrum, as illustrated in Figure 1(c).

## 3. RESULT AND DISCUSSION

### A. NUMERICAL SIMULATIONS

As an initial step, we performed a series of numerical simulations that emulate a THz Fourier Ptychographic imaging experiment, as illustrated in Figure 1(a). We numerically model the experiment as a close analogy to the realistic experimental scenario, where a CW source operating at a 3.5 THz frequency is employed to illuminate the test phase sample (see Figure 2(a)). Following the interaction with the sample, the transmitted intensity images are numerically recorded at the imaging plane within the FOV of 17.5 × 17.5 mm², discretized into a 500 × 500 pixel grid with a pixel pitch of 35 μm (similar to the THz camera used in the experiment). Additionally, to realistically assess the performance of our technique under the detector noise conditions comparable to the performance metrics of a commercially available microbolometric THz camera, we introduce 20 dB additive white Gaussian noise into the stack of emulated intensity data corresponding to multi-angle illuminations. In such a case, a recorded diffraction-limited intensity image corresponding to on-axis illumination is shown in Figure 2(b). Notably, due to the limited NA (0.2 in our simulation), the detection optics function as a low-pass spatial frequency filter, restricting the resolvable feature size to ~260 μm according to the Rayleigh criterion. This leads to the loss of fine structural details and, consequently, reduces the visibility of sub-diffraction features in the recorded intensity image. In contrast, our iterative Ptychographic phase retrieval approach reconstructs the high-resolution phase image of the target sample. As a relevant case example, the reconstructed phase image is shown in Figure 2(c), illustrating an improvement in spatial resolution and the accurate recovery of fine structural details. Figure 2(d) illustrates the quantitative insight into the enhancement of spatial resolution by comparing line-cuts in the phase image object ('a' red line in Figure 2(a)) and reconstructed phase image ('b' cyan line in Figure 2(c)), hence showcasing the efficiency of our approach in improving resolution. Quite interestingly, our imaging technique can fully resolve the 100-μm slit separated by 140 μm, present in the imaging object, by synthetically expanding the effective NA to over 0.5 through multi-angle illumination, albeit with a marginally reduced phase contrast that mainly depends on SNR and optical properties of the sample. To evaluate the convergence behavior of the iterative Ptychographic phase retrieval scheme, Figure 2(e) shows the variation of the structural similarity index measure (SSIM) [50] of the reconstructed phase image as a function of the number of iterations, demonstrating rapid convergence towards a stable solution, with an SSIM of 0.8 achieved in only 60 iterations. Furthermore, to analyze the robustness of our reconstruction routine in the presence of the noisy detection, we show the relationship between the SSIM of the reconstructed phase image and the SNR detection error in Figure 2(f). The result demonstrates that our Ptychographic phase retrieval technique could successfully achieve reliable reconstruction with an SNR threshold as low as 15 dB. Such resilience showcases the practical applicability and strength of our Ptychographic imaging framework under noisy measurement scenarios.

### B. EXPERIMENTAL SETTINGS

The principal elements of our experimental implementation of the THz Fourier Ptychographic imaging scheme are shown in the supplementary figure S1. We employ a CW QCL laser source (Lytid TeraCascade 2000 series) operating at a monochromatic frequency of 3.5 THz, emitting a diverging beam with an output power of ~ 13 mW (~ 90% of the source power). The beam collimation and spatial coherence enhancement of the source beam are achieved by using a collection of optical components comprising three off-axis parabolic mirrors of focal lengths 15.24 cm, 22.86 cm, and 7.62 cm, respectively, in combination with a pinhole. The resulting collimated and spatially coherent beam is then directed onto a motorized mirror (Thorlabs Z912 motorized kinematic mount) that introduces a controlled tilt in the beam wavefront. The tilted wavefront is subsequently projected onto the imaging sample through a 4f lens configuration consisting of two off-axis parabolic mirrors, each with a focal length of 15.24 cm. This arrangement ensures that the motorized mirror plane is optically conjugate to the sample plane, thus enabling precise translation of tip and tilt in the incident wavefront onto the sample. Conversely, the transmitted field after interaction with the sample is imaged through a 4f telescopic lens system (focal length 3.5 cm, each) onto a microbolometric THz camera (INO MICROXCAM-384i), which offers a 384 × 288 pixels intensity image with a pixel pitch of 35 μm.

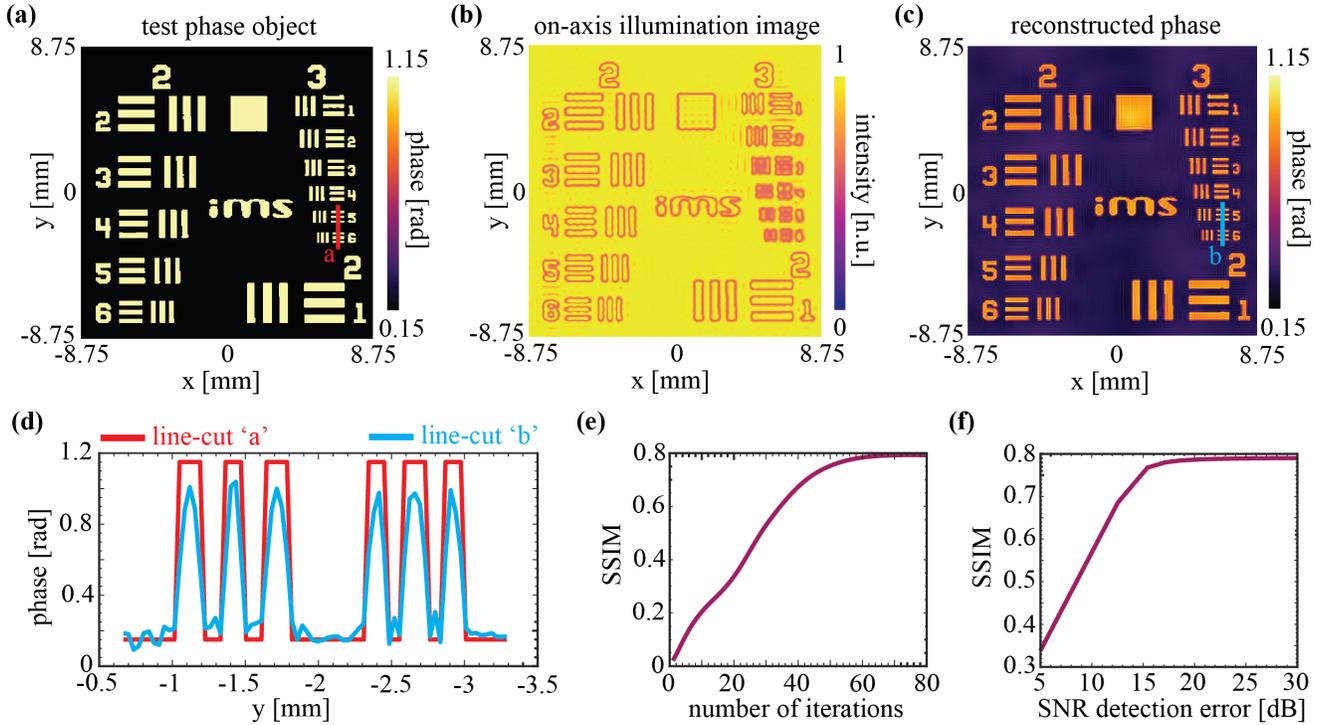

**Figure 2: Numerical analysis of THz Fourier Ptychographic imaging scheme. (a)** Target phase imaging object as a ground truth. **(b)** diffraction-limited image obtained under on-axis illumination. **(c)** Reconstructed high-resolution phase image. **(d)** A line-cut comparison between the target phase object (red) and the reconstructed phase image (cyan). **(e)** Variation of the phase SSIM with the number of iterations. **(f)** Variation of phase SSIM as a function of signal-to-noise ratio (SNR) detection error. The 17.5 × 17.5 mm² illumination area is spatially sampled at 35 μm intervals, resulting in a total sampling grid of 500 × 500 pixels.

## C. ILLUMINATION ANGLE ESTIMATION

To reconstruct a large SBP image from a stack of intensity measurements, our method requires precise knowledge of illumination beam characteristics, particularly, the $\vec{k}_n$ vector of plane wave incident on the sample. Although the motorized kinematic mirror introduces controllable tilts in the source beam, the mechanical system remains sensitive to experimental misalignment errors and model mismatches, which can lead to artifacts during iterative Ptychographic phase reconstruction (as described in the Methods section). Consequently, extensive system calibration is necessary to ensure alignment between the computational phase retrieval routine and the experimental setup. This calibration process can be time-consuming and labor-intensive, requiring considerable user expertise. To overcome this issue, we introduce an illumination-tilt calibration procedure that operates directly on the recorded low-resolution intensity images obtained under varying illumination angles. Figure 3 illustrates the implementation of the illumination-angle calibration strategy proposed in the THz Fourier Ptychographic phase imaging framework. In this approach, a stack of experimentally recorded intensity images (Figure 3(a)) acquired under 30 different illuminations is considered, where the plane wave $\vec{k}$-vector of the source beam follows a predefined spiral trajectory. Recalling equation 2, the spatial Fourier transform of the $n^{\text{th}}$ intensity image corresponds to an autocorrelation of the sample spectrum. This leads to an interference between the DC component of the sample spectrum and the pupil function, resulting in the formation of two separate circles symmetrically centered at $(k_{xn}, k_{yn})$ and $-(k_{xn}, k_{yn})$. An example of such an intensity image and its spatial Fourier transform, recorded for the 23$^{\text{rd}}$ illumination, is shown in Figure 3(b) and Figure 3(c), respectively. It means that the illumination plane wave $\vec{k}$-vector (or illumination calibration) can be determined by looking for the centers of the circles (as shown in Figure 3(c)). In this context, identifying the centers of circles within the amplitude of spatial frequency space represents the core task of image processing, which can be intuitively formulated as a circular edge detection problem.

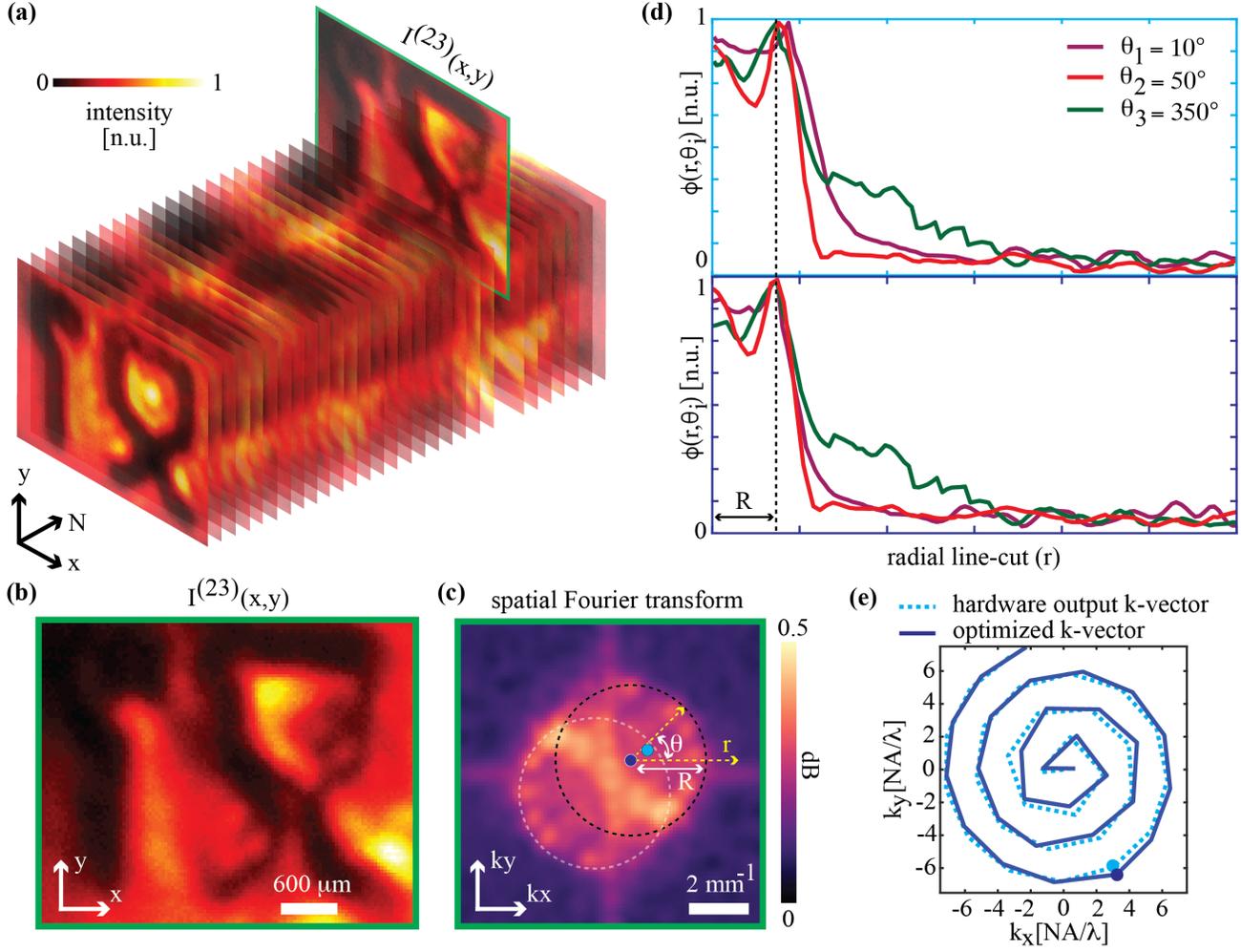

**Figure 3: Calibration of illumination angles in the proposed THz Fourier Ptychographic imaging.** (a) Recorded intensity stack of the test phase object under multiple illumination angles. (b) Recorded intensity pattern corresponding to the 23$^{rd}$ illumination angle. (c) Fourier-transformed image of (b); blue and cyan dots represent the calibrated and uncalibrated centers of the dashed reference circle (black), respectively. $(r, \theta)$ denotes the polar coordinate of the radial line with respect to the center, and R represents the calibrated radius. (d) Radial line-cuts corresponding to different $\theta$ values considering the uncalibrated (cyan box) and calibrated (blue box) points in (c). The line profiles signify the decent alignment of edge positions across all three angles considering the calibrated point whereas the edge positions are misaligned for the uncalibrated point (e) Hardware outputted (cyan dashed spiral path) and optimized (blue solid spiral path) illumination angle trajectories in the spatial frequency domain.

To this end, we assume $|\tilde{I}^{(n)}(k_x, k_y)|$ as a quasi-binary spatial distribution, assigning a value of 1 inside the two circles and 0 elsewhere. To approximate such behavior in the low SNR experimental data, we first normalize all the $N$ spectra by dividing them by the mean spectral amplitude, $\bar{\mu}(k_x, k_y) = \frac{1}{N}\sum_n |\tilde{I}^{(n)}(k_x, k_y)|$. The normalized data is then convolved with a Gaussian blur kernel to suppress high-frequency speckle noise. Figure 3(c) shows the smoothed spectrum corresponding to the 23$^{rd}$ illumination angle, where R denotes the circle radius and $(r, \theta)$ are the polar coordinates. Notably, if $\vec{k}'_n \equiv (k'_{xn}, k'_{yn})$ is the precise center of one of the circles, a sharp drop in the $|\tilde{I}^{(n)}(k_x, k_y)|$ should occur at R along any outward radial line $\phi(r, \theta_i)$ emanating from this center at any arbitrary angle $\theta_i$. Based on this principle, we begin with an initial estimate of the circle center and iteratively refine it by examining the radial profiles $\phi(r, \theta_i)$ until alignment of the edge positions across all the $\theta_i$ values are achieved. Figure 3(d) demonstrates this concept. The radial line profiles $\phi(r, \theta_i)$ are plotted for both the hardware-defined illumination vector $\vec{k}_n \equiv (k_{xn}, k_{yn})$ from the tilting mirror's positional output and the optimised vector $\vec{k}'_n \equiv (k'_{xn}, k'_{yn})$ determined through the circular edge detection scheme, evaluated at three angles ($\theta_1 = 10°, \theta_2 = 50°$, and $\theta_3 = 350°$). As shown, for the uncalibrated case (cyan box), the sharp intensity drops occur at inconsistent radial positions, whereas for the calibrated case (blue box), the edge positions are well aligned across all three angles, confirming accurate center estimation. This procedure is repeated for each recorded illumination condition, yielding calibrated incident wave vectors for all sequential illuminations. A comparison of the spatial trajectories of the incident $\vec{k}$ vectors obtained from the hardware outputs (uncalibrated) and those estimated from the circular edge detection routine are shown in Figure 3(e). Additionally, the reconstructed amplitude and phase distributions obtained using both the uncalibrated and calibrated illumination wave vectors are shown in Supplementary Figure S2.

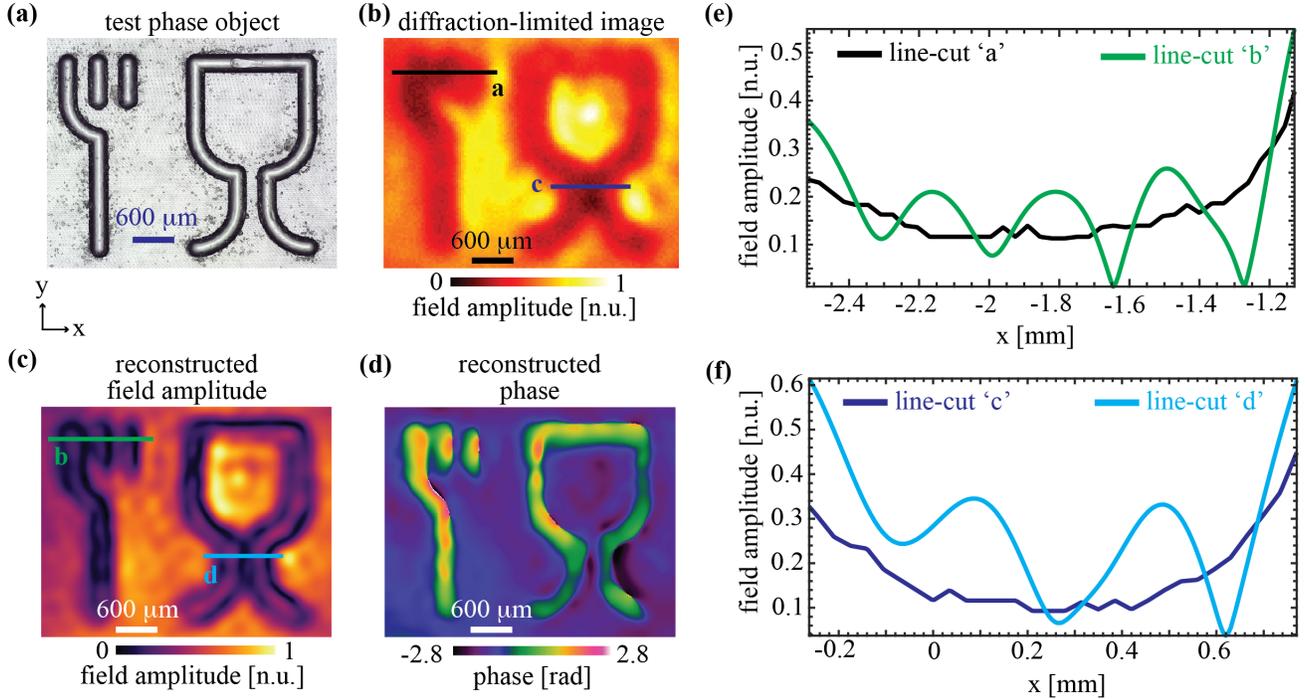

**Figure 4: Experimental results of THz Fourier Ptychographic imaging. (a)** Optical image of the test object under experimental investigation. **(b)** recorded diffraction-limited low-resolution amplitude image corresponding to on-axis illumination. **(c, d)** Reconstructed amplitude and phase images. **(e)** Line profile comparison across the "fork" region in the diffraction-limited (red line) and reconstructed (green line) images. **(f)** Line profile comparison across the "glass" region in the diffraction-limited (purple line) and reconstructed (cyan line) images. The 4.1 × 3.2 mm² illumination region is spatially sampled at 35 μm intervals, resulting in a total sampling grid of 117 × 91 pixels.

### D. SPATIAL RESOLUTION ENHANCEMENT BY THZ FOURIER PTYCHOGRAPHY

As a first case study, we employed a transparent plastic coffee stirrer as the imaging sample, focusing on the regions corresponding to the 'fork' and 'glass' features (Figure 4(a)). In this context, we record 30 diffraction-limited intensity images of the sample under varying illumination wavefronts impinging at various angles following an illumination k-vector spiral trajectory. In post-processing, the acquired intensity images are down-sampled to a 117 × 91-pixel region of interest, corresponding to a FOV of 4.1 × 3.2 mm² with a pixel spacing of 35 μm. Figure 4(b) shows a diffraction-limited amplitude image of the target sample corresponding to an on-axis illumination, resulting in degraded resolution and structural features, particularly in the three sticks on the 'fork' and tail of the 'glass'. To recover high-resolution amplitude and phase information, we applied an iterative Ptychographic phase retrieval routine utilizing calibrated illumination information. The reconstructed amplitude and phase images are shown in Figures 4(c) and 4(d), respectively, and reveal a substantial improvement in spatial resolution, as well as the faithful recovery of fine structural details that are otherwise unresolved in conventional diffraction-limited THz imaging.

A quantitative comparison between the diffraction-limited and reconstructed results is presented in Figures 4e and 4f, where we extract line profiles across regions of interest. Specifically, three sticks of the "fork" (each ~230 μm wide with ~150 μm spacing) in the diffraction-limited amplitude image, appears flat response (as shown in line profile 'a'), indicating an inability to resolve the fine structures. Similarly, the legs of the "glass" (separated by ~150 μm) remain unresolved in line profile 'c'. In contrast, the corresponding profiles 'b' and 'd' taken from the reconstructed amplitude image shows distinct peaks and valleys, demonstrating a substantial enhancement in spatial resolution. For reference, an optical image acquired with an advanced microscope is shown in Supplementary Figure S3. Notably, our proposed THz Fourier Ptychographic imaging approach expands the accessible spatial frequency bandwidth by a factor of 6.5 relative to the diffraction-limited measurement, corresponding to a 2.5× increase in the synthetic NA of the imaging system.

To demonstrate the practical applicability of our proposed imaging framework, we further examined physical features and texture geometry in the target imaging samples. As an illustrative case, we employed a transparent polypropylene (PP) plastic disk featuring an embossed "1" pattern (Figure 5(a)) to evaluate its surface topography. We record 30 diffraction-limited intensity images corresponding to various impinging wavefronts, with the k-vector of the illumination following the spiral trajectory. The recorded intensity data is then processed using the Fourier Ptychographic iterative phase retrieval routine to reconstruct the spatial phase distribution of the sample. Once the spatial phase map ($\Delta\phi(x,y)$) is obtained, the surface topographic variations in the sample can be directly estimated as, $\Delta h(x,y) = \frac{\lambda}{2\pi} \frac{\Delta\phi(x,y)}{(n_m - n_0)}$, where $n_m = 1.51$ refractive index of PP material [51] and $n_0 = 1$ corresponds to the refractive index of the air. The resulting quantitative surface topographic map is presented in Figure 5(b). Remarkably, the reconstructed height profile reveals fine structural details that closely correspond to the embossed features observed in the optical profilometry measurements (Supplementary Figure S4), thus validating the accuracy and reliability of our approach for quantitative surface characterization.

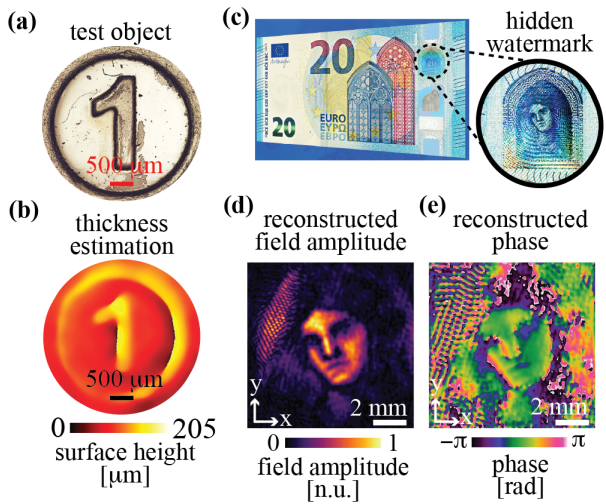

**Figure 5: THz phase imaging for material characterization and quality control. (a)** Optical image of the polypropylene (PP) plastic phase object. **(b)** Estimated surface topographic map of the object. **(c)** Optical image of €20 banknote showing the hidden watermark of '*Europa*'. **(d, e)** Reconstructed amplitude and phase images of the watermark. The 10.1 × 10.1 mm² illumination region is spatially sampled at 35 μm intervals, resulting in a total sampling grid of 288 × 288 pixels.

As a final demonstration, we show the ability of our THz Fourier Ptychographic imaging approach to recover concealed subsurface structures, demonstrating its potential for currency authentication and counterfeit detection. In this context, we use a 20€ banknote containing a subsurface security feature known as 'portrait window', which conceals a watermark of '*Europa*', visible only when the banknote is held against the light (Figure 5(c)). We recorded 30 diffraction-limited intensity images, each of 288 × 288 pixels, covering a field of view of 10.1 × 10.1 mm² with a pixel spacing of 35 μm, corresponding to the multi-angle illumination. From the acquired information, we reconstructed the amplitude and phase images of the hidden watermark, as illustrated in Figures 5(d) and 5(e), respectively, illustrating efficient recovery of subsurface structural variations and security features embedded in banknotes.

In terms of the experimental timeframe, 30 images corresponding to 30 different illuminations are acquired within 30s (integration time: 1s per frame) using an automated LabVIEW-controlled program. The iterative reconstruction is performed on a Dell Precision 5480 equipped with a 13th Gen Intel® Core™ i9-13900H processor (2.60 GHz) and 32 GB RAM, requiring approximately 90s to complete. This reconstruction time remains consistent across all test samples used in the experiments.

## 4. CONCLUSION

We have theoretically and experimentally demonstrated a THz Fourier Ptychographic imaging framework that surpasses the diffraction-limited resolution of conventional THz imaging systems by synthetically extending the effective NA. Our approach combines multi-angle plane-wave illumination with an aberration-corrected iterative phase-retrieval algorithm to recover high-resolution complex fields, resulting in quantitative amplitude and phase images with significantly expanded spatial-frequency bandwidth. Our imaging technique also incorporates an object-independent illumination-angle calibration strategy, based on circular edge detection in Fourier space, which enables the accurate estimation of illumination wave vectors and substantially suppresses reconstruction artifacts arising from geometric misalignment, thermal drift, and mechanical instabilities. By jointly refining the pupil function during reconstruction, our proposed Ptychographic phase retrieval routine compensates system-induced aberrations and improves convergence and phase reconstruction fidelity. Our THz imaging framework enables the quantitative spatial phase reconstruction of semi-transparent samples, achieving a 6.5× expansion of the recoverable spatial-frequency bandwidth, which corresponds to a 2.5× enhancement in the synthetic NA. Beyond resolution gains, we demonstrate practical utility by retrieving surface topography and reconstructing hidden subsurface security features in a currency note, highlighting remarkable potential for counterfeit detection and quality assurance. Our work establishes a robust and hardware-agnostic route to high-resolution THz phase imaging. The ability to non-destructively retrieve sample details with enhanced resolution establishes THz Fourier Ptychography as a valuable tool for quantitative inspection in material characterization, quality control, and security-sensitive authentication tasks.

**Funding.** Agence Nationale de la Recherche (ANR-22-CE42-0005-HYPSTER, ANR 22-PEEL-0003-Comptera).

**Author Contributions.** All authors were engaged in the discussion regarding the basic concept of the paper and its implementation for THz waves. P.Mu. performed the experiment and drafted the initial manuscript, with contribution from all authors on the interpretation of the results. V.K. and P. Mo. supervised the project.

**Disclosures.** The authors declare no conflicts of interest.

**Data availability.** The datasets for all figures are freely accessible at *<link>*

**Supplemental document**. Supplementary figures can be found here.

*Supplementary Information*

# Terahertz Fourier Ptychographic Imaging

**Pitambar Mukherjee[1], Vivek Kumar[1,2], Frederic Fauquet[1], Amaury Badon[3], Damien Bigourd[1], Kedar Khare[4], Sylvain Gigan[2], Patrick Mounaix[1]**

[1] IMS Laboratory, University of Bordeaux, UMR CNRS 5218, 351 Cours de la Libération Bâtiment A31, 33405 Talence, France.
[2] Laboratoire Kastler Brossel, ENS-Université PSL, CNRS, Sorbonne Université, Collège de France, 24 rue Lhomond, 75005 Paris, France.
[3] Laboratoire Photonique Numérique et Nanosciences (LP2N), UMR 5298, University of Bordeaux, 33400 Talence, France.
[4] Optics and Photonics Centre, Indian Institute of Technology Delhi, New Delhi 110016, India.

*The supplementary information consists of 4 figures.*

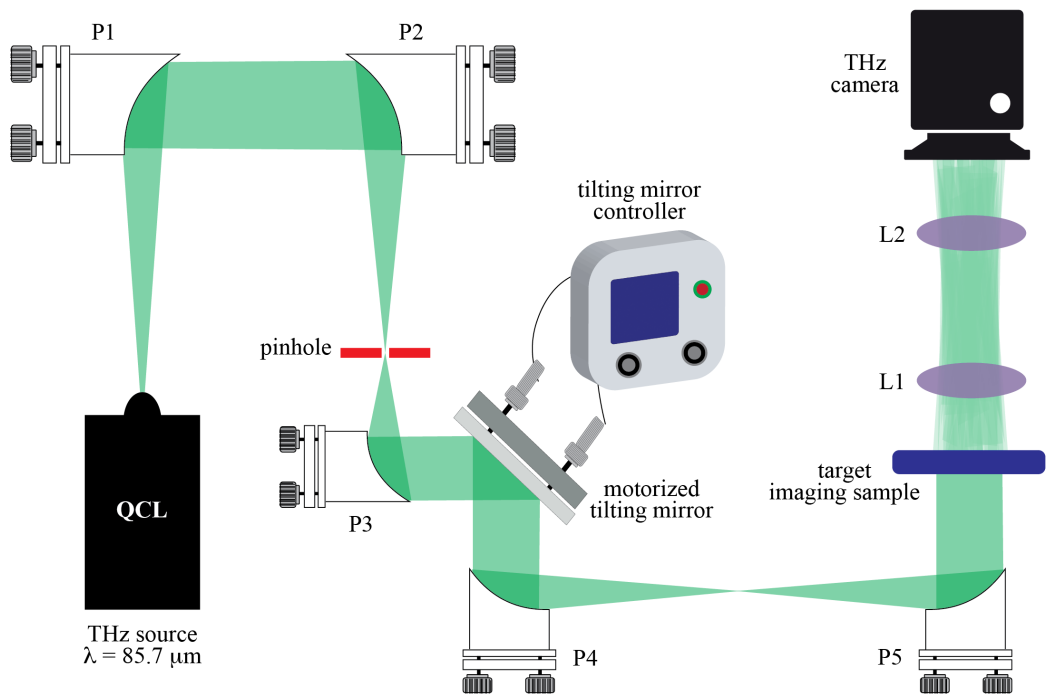

**Figure S1: Detailed experimental schematic of THz Fourier ptychographic setup.** P1, P2, P3, P4 and P5: off-axis parabolic mirrors; L1 and L2: TPX aspheric plano-convex lenses forming a 4f imaging system.

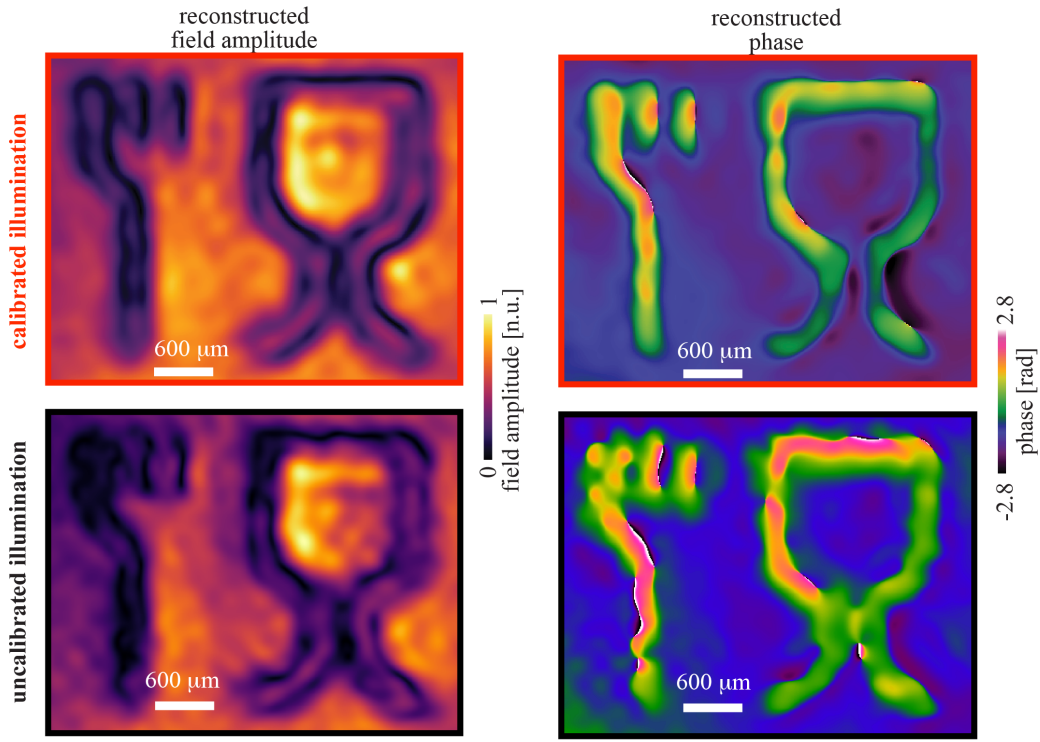

**Figure S2:** Reconstructed amplitude and phase of the imaging sample associated with the calibrated and uncalibrated illumination wave vectors, respectively.

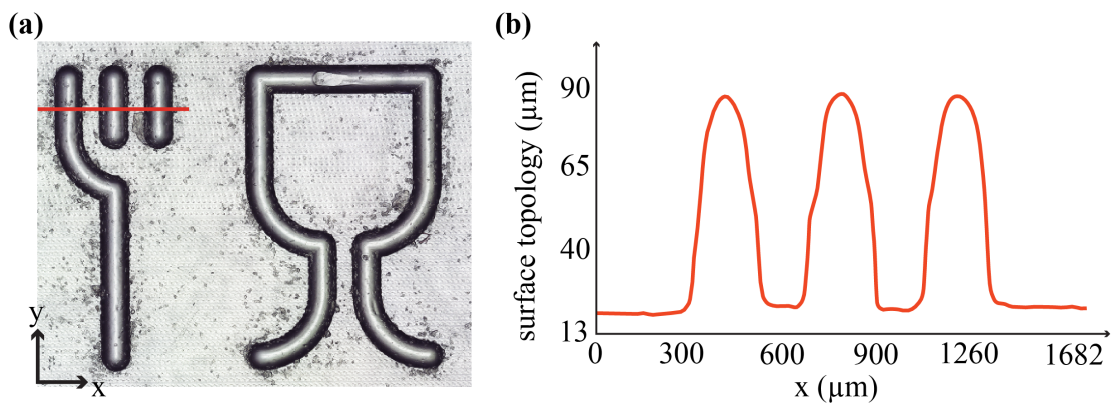

**Figure S3: Optical characterization of the transparent plastic coffee stirrer sample. (a)** Optical image captured using an advanced optical microscope. The image highlights the 'fork' region of the stirrer, which consists of three distinct prongs. **(b)** Line profile analysis obtained along the three prongs in the fork section.

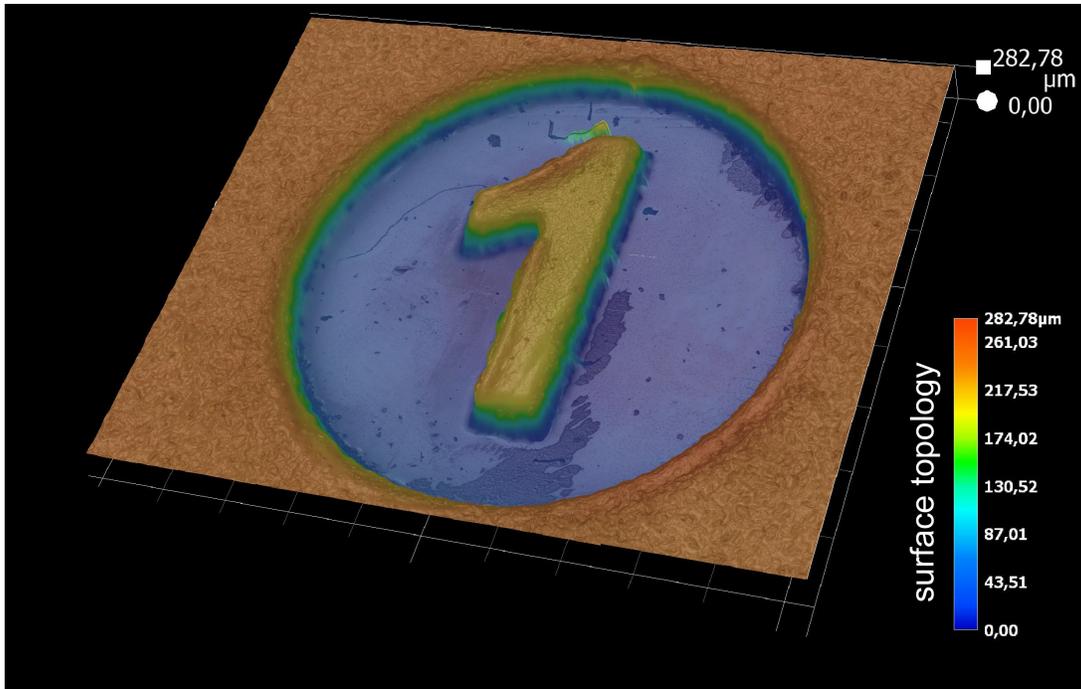

**Figure S4:** Optical topographic profile of the polypropylene (PP) plastic "1" sample captured using an advanced optical microscope, showing the surface features and morphology.